\documentclass[fleqn,10pt]{wlscirep}

\usepackage[utf8]{inputenc}
\usepackage[T1]{fontenc}
\usepackage{makecell}

\makeatletter
\newcommand\thefontsize{The current font size is: \f@size pt}
\makeatother

\usepackage{subcaption}

\usepackage{gensymb}
 \usepackage[acronyms]{glossaries}	
\newacronym{CXR}{CXR}{chest x-ray}
\newacronym{CAD}{CAD}{computer-aided diagnosis}
\newacronym{IoU}{IoU}{intersection over union}
\newacronym{CNN}{CNN}{convolutional neural network}
\newacronym{NCC}{NCC}{normalized cross-correlation}
\newacronym{IRB}{IRB}{Institutional Review Board}

\usepackage{verbatim}

\title{REFLACX, a dataset of reports and eye-tracking data for localization of abnormalities in chest x-rays}

\author[1,*]{Ricardo Bigolin Lanfredi}
\author[2]{Mingyuan Zhang}
\author[3]{William F. Auffermann}
\author[3]{Jessica Chan}
\author[3]{Phuong-Anh T. Duong}
\author[4]{Vivek Srikumar}
\author[5]{Trafton Drew}
\author[3]{Joyce D. Schroeder}
\author[1]{Tolga Tasdizen}
\affil[1]{Scientific Computing and Imaging Institute, University of Utah, 72 S Central Campus Drive, Room 3750, Salt Lake City, UT 84112, USA}
\affil[2]{Department of Population Health Sciences, University of Utah, 295 Chipeta Way, Williams Building, Room 1N410,
Salt Lake City, UT 84108, USA}
\affil[3]{Department of Radiology and Imaging Sciences, University of Utah, 30 North 1900 East \#1A071, Salt Lake City, UT 84132, USA}
\affil[4]{School of Computing, University of Utah, Room 3190, 50 Central Campus Dr., Salt Lake City, UT 84112, USA}
\affil[5]{Department of Psychology, University of Utah, 380 S 1530 E Beh S 502, Salt Lake City, UT 84112, USA}
\affil[*]{corresponding author(s): Ricardo Bigolin Lanfredi (ricbl@sci.utah.edu)}

\begin{abstract}
Deep learning has shown recent success in classifying anomalies in chest x-rays, but datasets are still small compared to natural image datasets. Supervision of abnormality localization has been shown to improve trained models, partially compensating for dataset sizes. However, explicitly labeling these anomalies requires an expert and is very time-consuming. We propose a potentially scalable method for collecting implicit localization data using an eye tracker to capture gaze locations and a microphone to capture a dictation of a report, imitating the setup of a reading room. The resulting REFLACX (Reports and Eye-Tracking Data for Localization of Abnormalities in Chest X-rays) dataset was labeled across five radiologists and contains 3,032 synchronized sets of eye-tracking data and timestamped report transcriptions for 2,616 chest x-rays from the MIMIC-CXR dataset. We also provide auxiliary annotations, including bounding boxes around lungs and heart and validation labels consisting of ellipses localizing abnormalities and image-level labels. Furthermore, a small subset of the data contains readings from all radiologists, allowing for the calculation of inter-rater scores.
\end{abstract}
\begin{document}

\flushbottom
\maketitle

\thispagestyle{empty}

\section*{Background \& Summary} 
Deep learning has been successfully applied to medical image analysis, including abnormality detection in \glspl*{CXR}~\cite{chexnet,radreview}. However, even though, in the medical image context, very large published \gls*{CXR} datasets are available~\cite{padchest,chexpert,chestxray8,mimiccxr}, with sizes in the hundreds of thousands of images, they are much smaller than some natural images datasets~\cite{imagenet,tencentdataset}. These large \gls*{CXR} datasets were labeled through the mining of clinical reports. However, given the relatively small size of the datasets, additional labels may contribute to the training of data-demanding deep convolutional neural networks. We present a dataset named REFLACX containing eye-tracking data collected from radiologists while they dictate reports. This dataset was built as a proof-of-concept for a data collection method that expands the labels of a medical dataset, providing additional supervision during deep learning training. 

Li et al. have shown that bounding boxes localizing abnormalities on \glspl*{CXR} can be used to supervise a \gls*{CNN} to improve accuracy and localization scores~\cite{cvpr2018}. They used the manually annotated bounding boxes provided for 880 images of the 112,200 \glspl*{CXR} from the ChestX-ray14 dataset~\cite{chestxray8}. Even though other datasets provide similar labels~\cite{pneumoniadataset,vindrcxr}, such localization labels are rare, and when present, are usually provided in limited quantities, showing the difficulty of scaling up the manual labeling. To try to solve this issue, and in accordance with the prioritization of the development of automated image labeling and annotation methods of the NIH's roadmap for AI in medical imaging~\cite{roadmap}, we collected eye-tracking data from radiologists for implicit localization of anomalies. We believe that the proposed collection method has the potential to be scaled up and used as a nonintrusive annotation approach deployed in a radiology reading room.

Other works have used eye-tracking data to support models in training or inference. Templier et al. and Stember et al. used eye tracking for interactive segmentation of biological/medical images, intending to achieve faster labeling~\cite{segmic,segtumor}. Khosravan et al. proposed a method of integrating an eye tracker into the reading room to combine a radiologist's reading with \gls*{CAD} systems~\cite{falsenegativect}. Gecer et al. used the navigation behavior of pathologists during readings to improve the detection of cancer in histopathology images~\cite{pathologyzoom}. Stember et al. showed that the gaze of a radiologist when dictating standardized sentences for presence/absence of brain tumors in MRIs could be used to localize lesions~\cite{mridictation}. In parallel to our project, Karargyris et al. built a dataset of eye-tracking data and respective dictation of \gls*{CXR} reports, containing 1,083 readings by only one radiologist~\cite{ibm}. Saab et al. built a dataset of eye-tracking data for the task of identifying pneumothoraces for 1,170 \glspl*{CXR}~\cite{miccai2021}. Still, their dataset does not contain dictations.

\begin{figure}[ht]
\centering
\includegraphics[width=\textwidth]{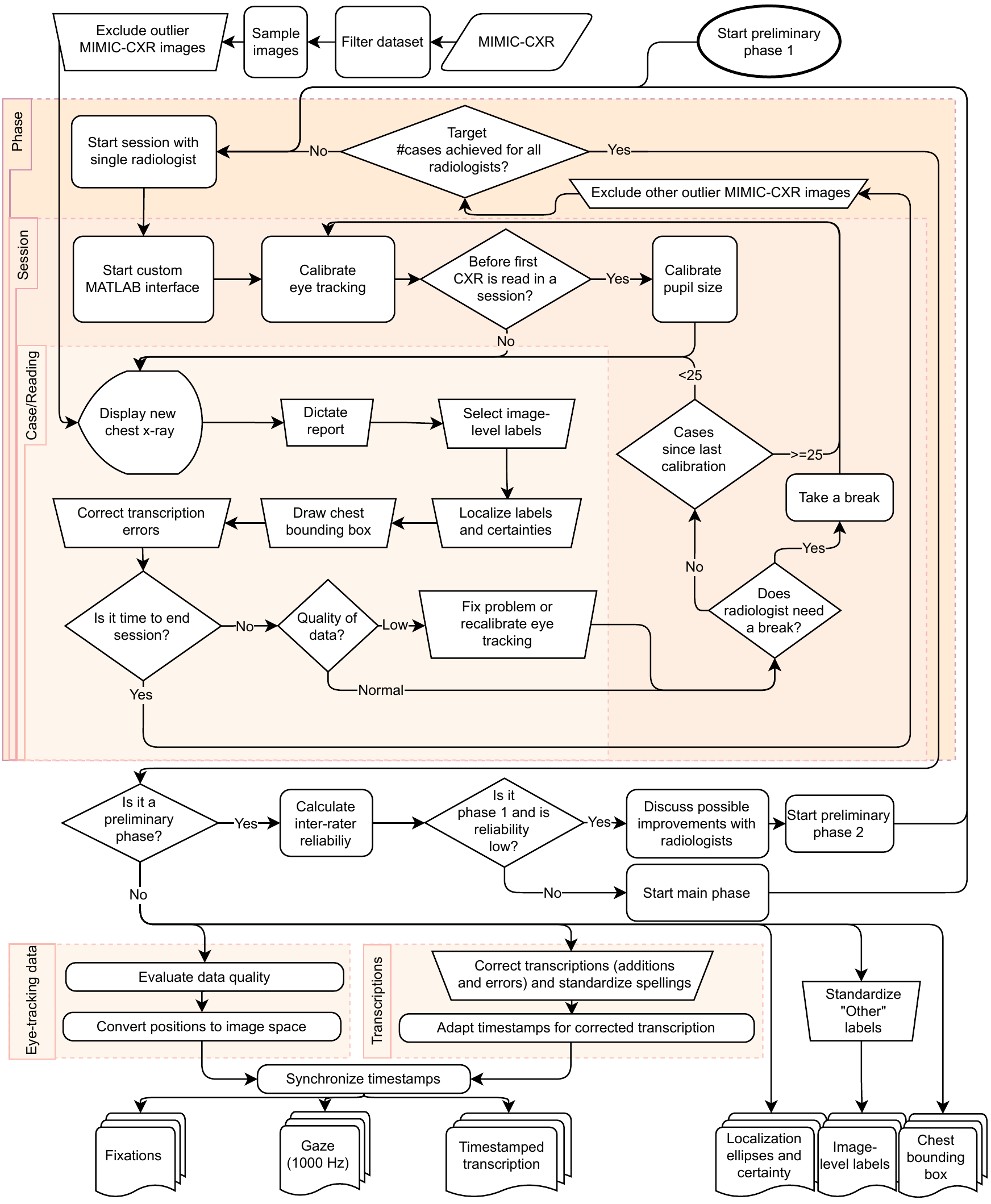}
\caption{Overview of the steps in the building of the dataset.}
\label{fig:overview}
\end{figure}

To collect the proposed dataset, as described in Figure~\ref{fig:overview}, five board-certified subspecialty-trained thoracic radiologists, who closely worked on aspects of the study design, used a custom-built user interface mimicking some of the functionalities from clinical practice. They dictated reports for images sampled from the MIMIC-CXR dataset~\cite{mimiccxr} while eye-tracking data were collected, including gaze location and pupil area. The audio of the dictation was automatically transcribed and manually corrected. The transcriptions included word timestamps for synchronization with the rest of the data. During dictation, we also recorded the zooming, panning, and windowing state of the \gls*{CXR} at all times. After dictating each case, radiologists provided a set of manual labels: the abnormalities they found, selected from a list, and the location of those abnormalities. These manual labels were collected to validate the data collection method and any automatically extracted labels generated from the dataset. They are not easily scalable and would not be collected in more extensive implementations of the collection method. Radiologists also drew a bounding box around the heart and lungs for normalization of chest position.

The collection of data was separated into three phases. The two first preliminary phases were used to adjust minor data collection details and to estimate inter-rater scores for the labels of the dataset, with radiologists reading a shared set of images. For the third and main phase, the sets of \glspl*{CXR} for each radiologist were independent. Considering all phases, REFLACX contains 3,052 cases, for a total of 2,616 unique \glspl*{CXR}. Of the 3,052 cases, 3,032 contain eye-tracking data.
\section*{Methods}
Data collection was divided into three phases. Phases 1 and 2 were preliminary phases where five radiologists read the same set of 59 and 50 \glspl*{CXR}, respectively. The set of possible image-level labels was chosen after a discussion among radiologists. After estimating the inter-rater reliability for phase 1, as shown in Table \ref{tab:fleiss}, a meeting was organized where radiologists discussed the labeling differences for the five cases that had the most negative impact on the reliability scores. The set of labels for phase 2 was slightly modified to clarify labeling and reduce its complexity. An electronic document, included as Supplementary File 1, was distributed to all radiologists with labeling instructions. A glossary by Hansell et al. provided some of the labeling definitions for this document~\cite{definitionopacification}. Phase 3 had the same five radiologists reading independent sets of around 500 \glspl*{CXR} each and used the same set of anomaly labels as phase 2. This phase constitutes the main content of the dataset and has a slightly higher quality of eye-tracking data. The set of labels used for each phase is listed in Table \ref{tab:stats}. Phase 1 went from November 11, 2020 to January 4, 2021, phase 2 from March 1, 2021 to March 11, 2021, and phase 3 from March 24, 2021 to June 7, 2021. Data collection sessions took on average 2.21 hours, with a maximum of 3.92 hours and a minimum of 0.2 hours.

This study complied with all relevant ethical regulations. Eye-tracking data collection was exempted from approval by the University of Utah \gls*{IRB}, and no informed consent was necessary. The use of the MIMIC-CXR \glspl*{CXR} was exempted from approval for this study since the images came from a publicly available and de-identified dataset. The MIMIC-CXR dataset was originally approved by the \gls*{IRB} of BIDMC, and patient consent was waived.

\subsection*{Image data}

\Glspl*{CXR} were sampled from the MIMIC-CXR dataset~\cite{mimiccxr,mimiccxrdicom,physionet}, a publicly available deidentified dataset. Before a random sampling of the images shown to radiologists, the dataset was filtered for including only \glspl*{CXR} that contained ``ViewPosition'' metadata with value ``PA'' or ``AP'', were the only frontal \glspl*{CXR} in their study, and were present in the label table from the MIMIC-CXR-JPG dataset~\cite{mimiccxrjpg,mimiccxrarxiv,physionet}. Images were sampled to include 20\% of images from the test set of the MIMIC-CXR dataset and 80\% from the other splits, and uniformly at random from each of these two splits. After the sampling for phase 3, images with anonymizing black boxes that intersected the lungs were manually excluded before presentation to radiologists. After the sessions, we manually excluded outlier images that, according to the radiologists, had major parts of the lung missing from the field of view, were digitally horizontally flipped, or had a rotation of 90\degree.

\subsection*{Data-collection sessions}
Data collection for each \gls*{CXR} involved two main parts: a dictation and transcription of a free-form radiology report while collecting eye-tracking data, and the selection of labels and ellipses to use as evaluation ground-truth for anomalies present in a \gls*{CXR} and their location. The data collection interface was developed in MATLAB R2019a/Psychtoolbox 3.0.17~\cite{ptb1,ptb2,ptb3}. The code for the interface is available at \href{https://github.com/ricbl/eyetracking}{https://github.com/ricbl/eyetracking}~\cite{mycodezenodo}. The interface is shown in Figure \ref{fig:screens} and as a video in Supplementary File 2, where the moving semitransparent red ellipse, with an axis length of 1\degree\ of visual angle, represents stabilized gaze, i.e., fixations, and the moving blue ellipse represents the instantaneous gaze location sample. The cursor was not drawn in the video and the audio is a digitally generated voice representing the timestamped transcription. We did not include the original dictation in the video for anonymization purposes. This interface allowed for:

\begin{itemize}
\itemsep 0em 
\item   Dictation of reports. A \gls*{CXR} was displayed to radiologists, and they dictated a report using a handheld microphone while eye-tracking data were collected. Radiologists did not have previous access to the \glspl*{CXR} or their reports. Eye-tracking data were collected as soon as the \gls*{CXR} was shown to radiologists, including data containing a reasoning period when they first saw the image and a dictation period. Radiologists were instructed to dictate reports as they would dictate in clinical practice. We also asked them to dictate free-form reports, so radiologists that used templates in clinical practice had to change their style to free-form.
\item   Editing of mistakes of the automated transcription. Radiologists were instructed to not add or change report content, and only correct transcription mistakes.
\item	Selection of image-level labels and ellipses localizing each label, including assessing the certainty of the presence of anomaly for each ellipse. Following Panicek et al.~\cite{certainty}, the allowed certainties were: Unlikely (<10\%), Less Likely (~25\%), Possibly (~50\%), Suspicious for/Probably (~75\%), Consistent with (>90\%). Radiologists were instructed to add image-level labels even in case they forgot to mention them in the original report.
\item	Image windowing, zooming, and panning. These features were available while dictating and labeling, mimicking features available to radiologists in their clinical practice.
\item	Drawing of chest bounding boxes that encompass lungs and heart, for normalization of the variations in \glspl*{CXR}.
\item   Calibration of pupil size and access to the eye-tracker calibration screens.
\end{itemize}

In more detail, at the beginning of the session, the calibration interaction adopted the following order:
\begin{enumerate}
\item Eye-tracking calibration screen, where 13 circles, distributed throughout the screen, are displayed in random order for the radiologists to fixate.
\item Calibration of pupil size.
\end{enumerate}

For the rest of the session, for each reading of \gls*{CXR}, the interaction of radiologists followed:
\begin{enumerate}
\item Dictation of report. Radiologists were not allowed to return to this screen for changes in dictation.
\item Image-level label selection.
\item Text input for ``Other'' label, in case it was selected as an image-level label.
\item Localization of image-level labels by the drawing of ellipses and choice of certainty for each ellipse.
\item Drawing of a chest bounding box including lung and heart.
\item Text editing to correct transcription errors.
\item Screen allowing proceeding to the subsequent \gls*{CXR}, displaying warnings for low-quality eye-tracking data, and optional return to calibration screens.
\end{enumerate}
\subsubsection*{Equipment}
To collect eye-tracking data, we used an Eyelink 1000 Plus system, which allows for high spatial (less than 1\degree of visual angle) and temporal (1,000 Hz) resolution. It also allows the radiologists to move their heads within a small area while maintaining this high degree of spatial and temporal acuity. Given our interest in ensuring that we had high-quality eye-tracking data, the experimenter calibrated and validated each radiologist multiple times throughout their viewing sessions. This process can be time-consuming, particularly if the clinician wears bifocals, which often lead to poor calibration. One alternative to this setup would have been to use a mobile eye-tracking system. These systems typically have the eye-tracking apparatus embedded within glasses worn by the subject. A good calibration can be achieved much more easily and possibly without an experimenter guiding this process. However, at present, the resultant data are much more difficult and time-consuming to analyze because most mobile eye-tracking systems do not co-register eye movements with precise screen coordinates. In practice, this typically means that the experimenters must hand-code the co-registration from a video output. This process is time-consuming and can suffer from bias, requiring multiple coders to examine the data to ensure reliable coding. 

The Eyelink 1000 Plus was equipped with a 25 mm lens and managed by an Eyelink 1000 Host PC recording gaze at 1,000 Hz. The eye tracker was configured in the remote setup, for which the radiologists put a sticker on their forehead. Radiologists had the freedom to move their heads as long as the sticker and tracked eye stayed within the camera's field of view, and the sticker stayed between 55 cm and 65 cm from the camera. The camera was positioned below a 27 inches BenQ PD2700U, 3,840$\times$2,160 pixels, 60 Hz screen, connected to a Display PC running Ubuntu 18.04. For phase 3, the camera was positioned 11 cm in front of the screen. Eye distance to the bottom of the screen was around 71 cm, while it was around 65 cm to the top of the screen.

\subsubsection*{Calibration}

Eye-tracker calibrations are necessary for finding the correspondence between pupil and cornea positions in the image captured by the eye tracker's camera and locations on the screen where the radiologists are looking. During the calibration, radiologists had to look at 13 targets in several locations on the screen for the eye-tracker to register the pupil and cornea positions for each location and interpolate for other intermediate locations on the screen. We performed an eye-tracker calibration at the beginning of each session, every 25 cases, every time radiologists took a break, or when noticeable quality problems with the data could be solved with a new calibration. These quality problems mainly involved moments when the cornea or the pupil was not recognized for specific eye positions or when the eye-tracker gave incorrect locations for the pupil or the cornea, e.g., glasses recognized as the cornea. The experimenter identified them with access to real-time data-collection information on the Eyelink 1000 Host PC. During calibration, radiologists were positioned such that the forehead sticker was between 59 cm and 61 cm away from the camera. For phases 1 and 2, calibrations were not performed at a regular 25-case interval. For phase 3, the calibration was considered successful if the average error was less than 0.6\degree, and the maximum error was less than 1.5\degree. The left eye was tracked by default, and the right eye was only tracked when the left eye calibration was repeatedly faulty. At each session beginning, radiologists were asked to look at the center of the screen for 15 s for measuring a constant used to normalize the pupil size.

\subsection*{Report transcription}
We collected the dictation audio at 48,000 Hz using a handheld PowerMic II microphone. The audio was transcribed using the IBM Watson Speech to Text service, which provides timestamps for each transcribed word, with the ``en-US\_BroadbandModel'' model. Before phase 1, the service's custom language model was trained with sentences from the ``Findings'' and ``Impressions'' sections of the MIMIC-CXR reports, which were filtered to remove sentences that referenced other studies of a patient through the search of keywords. For phase 2 and phase 3, in addition to language training with the filtered MIMIC-CXR reports, models had language and acoustic training with the collected audio and corrected transcriptions from phase 1.
Audio files had silence trimmed from the start and end of the file to speed up transcriptions. Silence was detected using Otsu's thresholding over the average audio level (dBFS) of 500 ms chunks. Word timestamps  were adjusted for the trimming of the beginning of the audio. After the report transcription was received from the cloud service, radiologists could make minor changes to it. Several of the common mistakes of the cloud service were programmatically corrected before providing transcriptions to the radiologists. 

\subsection*{Postprocessing}
The eye-tracking gaze samples were parsed for fixations, i.e., locations where gaze was spatially stabilized for a certain period; blinks, i.e., moments when the eye tracker did not capture pupil or cornea; and saccades, i.e., fast eye movements between fixations. Parsing was done in real time by the EyeLink 1000 Host PC, using a saccade velocity threshold of $35^{\circ}/s$, a saccade motion threshold of $0.2^{\circ}$, and a saccade acceleration threshold of $9,500^{\circ}/s^2$. Fixation locations were converted from screen space to image space by recording, at the start of the fixation, what image part was shown at what screen section. Fixations were synchronized with the transcriptions and other recorded data by synchronization messages sent by the Display PC to the EyeLink 1000 Host PC.

Pupil area data, captured by the eye tracker, was normalized by the average value of pupil area from the calibration screen from the beginning of the session. Radiologists were asked to look directly at the center of the screen, marked by a cross. The average value was weighted by fixation durations and calculated only for fixations at most 2\degree\ from the screen center.

After all data-collection sessions, transcriptions were checked and corrected by another person, who looked for additions of content during the radiologist correction screen, which were not allowed, and for out-of-context words and other clear mistakes, which were corrected by relistening to the recorded audios of those cases. During this process, spellings of a few words were standardized. The labels listed as ``Other'' were also standardized. Since the transcriptions were corrected for mistakes after receiving the output from the cloud service, timestamps had to be adapted to the new set of words. We used the counting of syllables to perform a linear interpolation between the times of both texts. Interpolations were calculated for each difference found between strings, as given by the difflib Python library. When there was an addition of words with no removal, the new words used the time range defined by the end time of the previous word and the start time of the next one, making it possible for words to have the same timestamp for start and end.

\subsubsection*{Data quality}
Readings were evaluated for the quality of eye-tracking data by measuring the times that data were classified as a blink during collection. Eye-tracking data were discarded in cases with a blink longer than 3 s or when blinks corresponded to more than 15\% of the data. Warnings were shown between cases for blinks longer than 1.5 s or when blinks corresponded to more than 10\% of the data. Cases that had eye-tracking data discarded were not included in the dataset for phase 3 but were included for phases 1 and 2 to make possible the evaluation of inter-rater scores for other labels.  The threshold values were chosen by qualitative observation of blink data histograms from phases 1 and 2 before collecting phase 3 data. Eye-tracking data were also discarded when the radiologist unintentionally clicked the ``Next Screen'' button before completing the dictation, when the eye-tracking data were not correctly saved because of various software problems, and when the eye tracker identified the glasses of the radiologist as their eyes. We discarded eye-tracking data for 7 cases for incomplete dictation, 6 for software problems when saving the data, 2 for large parts of the lungs missing, 1 for a horizontally flipped image,  1 for extreme rotation of the MIMIC-CXR image, 41 for low data quality, and 2 for glasses being confused for eyes. The total of discarded eye-tracking data was 10 cases for phase 1, 10 cases for phase 2, and 40 cases for phase 3. Additionally, one \gls*{CXR} in phase 1 had large parts of the lung missing and is not included in the 59 images present in the dataset. The 2,616 unique \glspl*{CXR} for which data are provided in the REFLACX dataset were included after manual image quality exclusions and data quality exclusions.

\section*{Data Records}
For each reading of a MIMIC-CXR image, the labels of this dataset consist of eye-tracking data, formatted as fixations and as gaze position samples, a timestamped report transcription, ellipses localizing anomalies associated with a certainty, and a chest bounding box. For each case/reading, there is a subfolder containing these labels, separated into individual tables. Subfolders from all three phases are grouped in the same folder (main\_data/ and gaze\_data/), and the phase to which each subfolder belongs is listed in metadata tables, one for each data collection phase. The statistics of the resulting dataset are presented in Table \ref{tab:stats}. Genotypical sex information was extracted from the MIMIC-IV dataset~\cite{mimiciv,physionet}, and was missing for around 0.35\% of the 2,199 unique subjects. The dataset is available on Physionet, at \url{https://doi.org/10.13026/e0dj-8498}~\cite{mydataset,physionet}. 

\subsection*{Description of tables and their columns}
\begin{itemize}
\item \textbf{main\_data.zip/main\_data/metadata\_phase\_<phase>.csv}: list of all the subfolders/cases corresponding to a specific phase and their metadata.
\begin{itemize}
\item \textbf{id}: subfolder name and unique identifier for a reading of a specific \gls*{CXR} by a specific radiologist.
\item \textbf{split}: the split given by the MIMIC-CXR dataset for that specific image. The possible values are ``train'', ``validate'', ``test''. Images were sampled so that 20\% of the images were from the test set of the MIMIC-CXR dataset.
\item \textbf{eye\_tracking\_data\_discarded}: for phases 1 and 2, even when the eye-tracking data were discarded for low quality, the anomaly labels, localizing ellipses, and chest bounding box were collected and included in the dataset. This column is ``True'' when the eye-tracking data has been discarded, and ``False'' otherwise. Transcriptions are also not included for these cases. Such cases should not be used for analysis if eye-tracking data or transcription are required. For phase 3, no case with discarded eye-tracking data is included.
\item \textbf{image}: path to the DICOM file from the MIMIC-CXR dataset used in this reading, with the same folder structure as provided by that dataset. 
\item \textbf{dicom\_id}: unique identifier of the image that can be used to join tables with the metadata from MIMIC-CXR.
\item \textbf{subject\_id}: unique identifier for the patient of that study.
\item \textbf{image\_size\_x}: horizontal size of the \gls*{CXR} in pixels
\item \textbf{image\_size\_y}: vertical size of the \gls*{CXR} in pixels
\item \textbf{Other columns}: the rest of the columns represent the possible presence of anomaly evidence in the image, as selected by the radiologist. Most of these columns contain values between 0 and 5, representing a certainty of the presence of such anomaly, according to the scale: 
\begin{itemize}
\item 0: not selected by radiologist,
\item 1: Unlikely, 
\item 2: Less Likely, 
\item 3: Possibly, 
\item 4: Suspicious for/Probably, 
\item 5: Consistent with.
\end{itemize}
Certainties were associated with each localizing ellipse in the image, so each label's maximum certainty is reported. Radiologists were asked not to draw ellipses for the anomaly labels ``Support devices,'' ``Quality issue'' and ``Other,'' so there is no certainty associated with these labels.
\item \textbf{Support devices} and \textbf{Quality issue}: the presence of these labels is reported, using ``True'' or ``False.''
\item \textbf{Other}:  A list of the other anomalies reported by radiologists not included in the rest of the labels, separated by ``|.'' If empty, no other anomaly was reported.
\end{itemize}
\item \textbf{main\_data.zip/main\_data/<id>/fixations.csv}: 
eye-tracking data summarized as fixations and collected during the dictation of the report.
\begin{itemize}
\item \textbf{timestamp\_start\_fixation /   timestamp\_end\_fixation}: 
the time in seconds when the fixation started/ended, counting from the start of the case.
\item \textbf{average\_x\_position /  average\_y\_position}: 
average position for the fixation, given in pixels and in the image coordinate space, where $(0,0)$ is the top left corner.
\item \textbf{pupil\_area\_normalized}: 
pupil area, normalized by the calibration performed at the beginning of each session.
\item \textbf{window\_level / window\_width}: 
average values of the windowing used for the image during a fixation. 
\item \textbf{angular\_resolution\_x\_pixels\_per\_degree / angular\_resolution\_y\_pixels\_per\_degree}: 
number representing how many image pixels fit in 1\degree\ of visual angle for each axis (x or y). It is dependent on the position of the fixation and the zoom applied to the image.
\item \textbf{xmin\_shown\_from\_image /  ymin\_shown\_from\_image /  xmax\_shown\_from\_image /\\ymax\_shown\_from\_image}: 
bounding box given in image-space representing what part of the image was shown to the radiologist at the start of the fixation. The reading/case always started with the whole image being shown, but zooming and panning were allowed.
\item \textbf{xmin\_in\_screen\_coordinates /  ymin\_in\_screen\_coordinates /  xmax\_in\_screen\_coordinates /\\ymax\_in\_screen\_coordinates}: 
bounding box given in screen space representing where the part of the image was shown.
\end{itemize}
\item \textbf{gaze\_data.zip/gaze\_data/<id>/gaze.csv}: complete eye-tracking data during the dictation of the report, collected at 1,000 Hz. Even though this data are not necessary for accomplishing the main research goals of this dataset, these data are included for any other analyses that need the gaze location in more detail than provided by the fixations.csv table. Compared to the fixations.csv table, the \textbf{timestamp\_start\_fixation} and \textbf{timestamp\_end\_fixation} columns were replaced by the \textbf{timestamp\_sample} column. The remaining columns are the same in both tables, but they represent values when the eye tracker’s camera captured the gaze sample in gaze.csv. 
\begin{itemize}
\item \textbf{timestamp\_sample}: timestamps do not start at 0 because audio recording started before gaze recording.
\item \textbf{x\_position / y\_position / pupil\_area\_normalized / angular\_resolution\_x\_pixels\_per\_degree /\\angular\_resolution\_y\_pixels\_per\_degree}: these columns are empty for timestamps when the eye tracker could not find the radiologist's pupil or cornea, making it impossible to calculate gaze at that moment. These rows are usually associated with moments when radiologists blinked.
\end{itemize}
\item \textbf{main\_data.zip/main\_data/<id>/timestamps\_transcription.csv}: timestamped corrected transcriptions of the reports. Radiologists were allowed to delete parts of the report and to modify transcription errors but not to add content.
\begin{itemize}
\item \textbf{word}: the word that was spoken. Periods (.), commas (,) and slashes (/) occupy one row.
\item \textbf{timestamp\_start\_word/timestamp\_end\_word}: the time in seconds when the dictation of each word started\slash ended, counting from the start of the case.
\end{itemize}
\item \textbf{main\_data.zip/main\_data/<id>/transcription.txt}: the transcription in text form, without timestamps.
\item \textbf{main\_data.zip/main\_data/<id>/anomaly\_location\_ellipses.csv}: bounding ellipses drawn by radiologists for each label present in the image. Each label may appear in more than one ellipse, and each ellipse may contain more than one label. Radiologists were instructed to select more than one label for an ellipse when a single image finding may be evidence of one or another label. Each row of the table represents one ellipse.
\begin{itemize}
\item \textbf{xmin /  ymin /  xmax /  ymax}: coordinates representing the extreme points, in image pixels, of the full horizontal and vertical axes of the ellipse. Coordinate $(0,0)$ represents the top left corner of the image.
\item \textbf{certainty}: value from 0 to 5 representing a certainty of the finding presence, according to the same scale as in the metadata table.
\item \textbf{Other columns}: the rest of the columns have a Boolean value representing the presence of evidence for each anomaly label in the ellipse. Since radiologists were asked not to draw ellipses for ``Support devices'', ``Quality issue'', and ``Other,'' most of the rows have the value ``False'' for these labels. For all other labels present in the image, at least one ellipse is drawn.
\end{itemize}
\item \textbf{main\_data.zip/main\_data/<id>/chest\_bounding\_box.csv}: single-row table containing information for the bounding box drawn around the lungs and the heart.
\begin{itemize}
\item \textbf{xmin /  ymin /  xmax /  ymax}: coordinates representing the extreme points, in image pixels, of the bounding box. Coordinate $(0,0)$ represents the top left corner of the image.
\end{itemize}
\end{itemize}
\section*{Technical Validation}
For all reported technical validation values, we also report standard error and sample size when applicable. Standard errors were only reported for $n>10$. For measurements that had more than one score for the same \gls*{CXR}, e.g., \gls*{IoU} calculated for all pairs of radiologists, we averaged the scores for each \gls*{CXR} before calculating the final average. The sample size given is the number of independent \glspl*{CXR} involved in the calculation.
\subsection*{Eye-tracking data}
Considering the errors provided by each of the calibrations used in data collection, we calculated the average and maximum calibration errors for each phase. Phase 1 had an average calibration error of 0.43$\pm$0.02\degree\ (n=25) and a maximum error of 2.79\degree, whereas, for phase 2, it was 0.43$\pm$0.03\degree\ (n=13) and 1.09\degree, respectively. Phase 3 had an average error of 0.44$\pm$0.01\degree\ (n=128)\ and a maximum error of 1.5\degree.

\begin{figure}[ht]
\centering
\includegraphics[width=\textwidth]{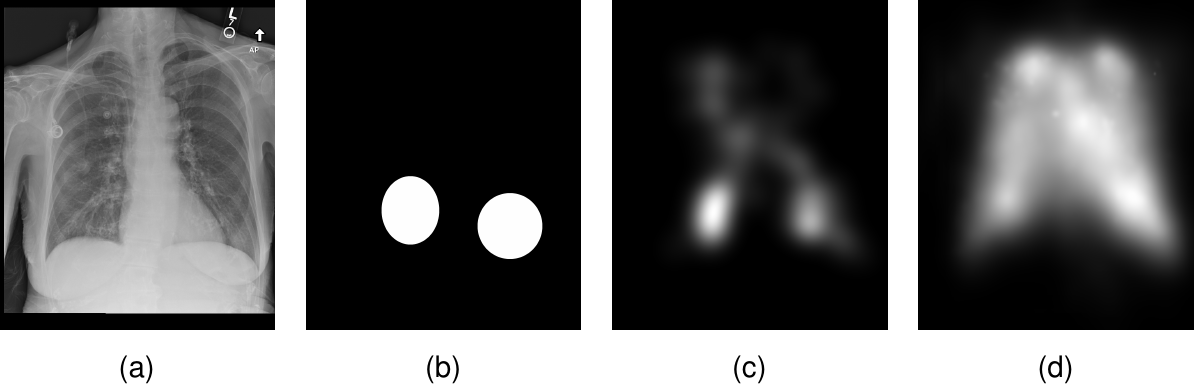}
\caption{Example of the localization information provided by the eye-tracking data and how it was validated. (a) CXR read by the radiologist. (b) Union of the abnormality ellipses selected by radiologists used to compare against heatmaps. (c) Heatmap generated by the fixations made by the radiologist while dictating the report. (d) Average heatmap for all radiologists and CXRs read in phases 1 and 2, normalized to the location of lung and heart of the CXR.}
\label{fig:heatmaps}
\end{figure}

To validate the presence of abnormality location information in the eye-tracking data, we calculated the presence of fixations inside the abnormality ellipses, as exemplified in Figure \ref{fig:heatmaps}. For each reading that had at least one drawn ellipse, we calculated the \gls*{NCC} between a fixation heatmap and a mask generated from the union of ellipses, using

\begin{equation}
\text{NCC}\left( x_f, x_e\right)=\frac{1}{P-1}\sum_{p }{\frac{\left({x_f}\left(p\right)-\mu_{x_f}\right)}{\sigma_{x_f}}\times\frac{\left(x_e\left(p\right)-\mu_{x_e}\right)}{ \sigma_{x_e}}},
\end{equation}

\noindent where $\text{NCC}\left( x_f, x_e\right)$ is the \gls*{NCC} between the fixation heatmap $x_f$ and the ellipse heatmap $x_e$, $\sigma_{x}$ is the standard deviation of a heatmap $x$, $\mu_x$ is the average value of a heatmap $x$, $x(p)$ is the value of a heatmap $x$ at pixel $p$, and $P$ is the number of pixels in the heatmaps. The fixation heatmaps were generated by drawing Gaussians centered on every fixation, with the standard deviation equal to 1\degree\ in each axis and with intensity proportional to the fixation duration. 

To check if the fixations heatmap of a specific \gls*{CXR} has more localization information than the heatmaps from unrelated \glspl*{CXR}, we compared the \gls*{NCC} for two types of fixations heatmap: heatmaps generated from the eye-tracking fixations of each \gls*{CXR} reading, as shown in Figure \ref{fig:heatmaps}c, and a baseline heatmap representing the average gaze over all \glspl*{CXR}, as shown in Figure \ref{fig:heatmaps}d. 

To calculate the baseline heatmap shown in Figure \ref{fig:heatmaps}d, we normalized all heatmaps to the same location using the labeled chest bounding boxes to compensate for the variation in the location of the lungs for each \gls*{CXR}. We calculated the average chest bounding box, transformed all heatmaps to this same space, and averaged the heatmaps. We finally transformed the average heatmap back to the space of each \gls*{CXR} to calculate the \gls*{NCC} against the labeled ellipses.

For the fixation heatmaps specific to each \gls*{CXR}, the average \gls*{NCC} achieved over all applicable \glspl*{CXR} was 0.380$\pm$0.014 (n=96), against a baseline score of 0.326$\pm$0.013 (n=96). This result shows that there is more abnormality location information in the fixations for each \gls*{CXR} than on a heatmap built from the usual areas looked at by a radiologist. 

\begin{figure}[ht]
\centering
\includegraphics[width=1.\textwidth]{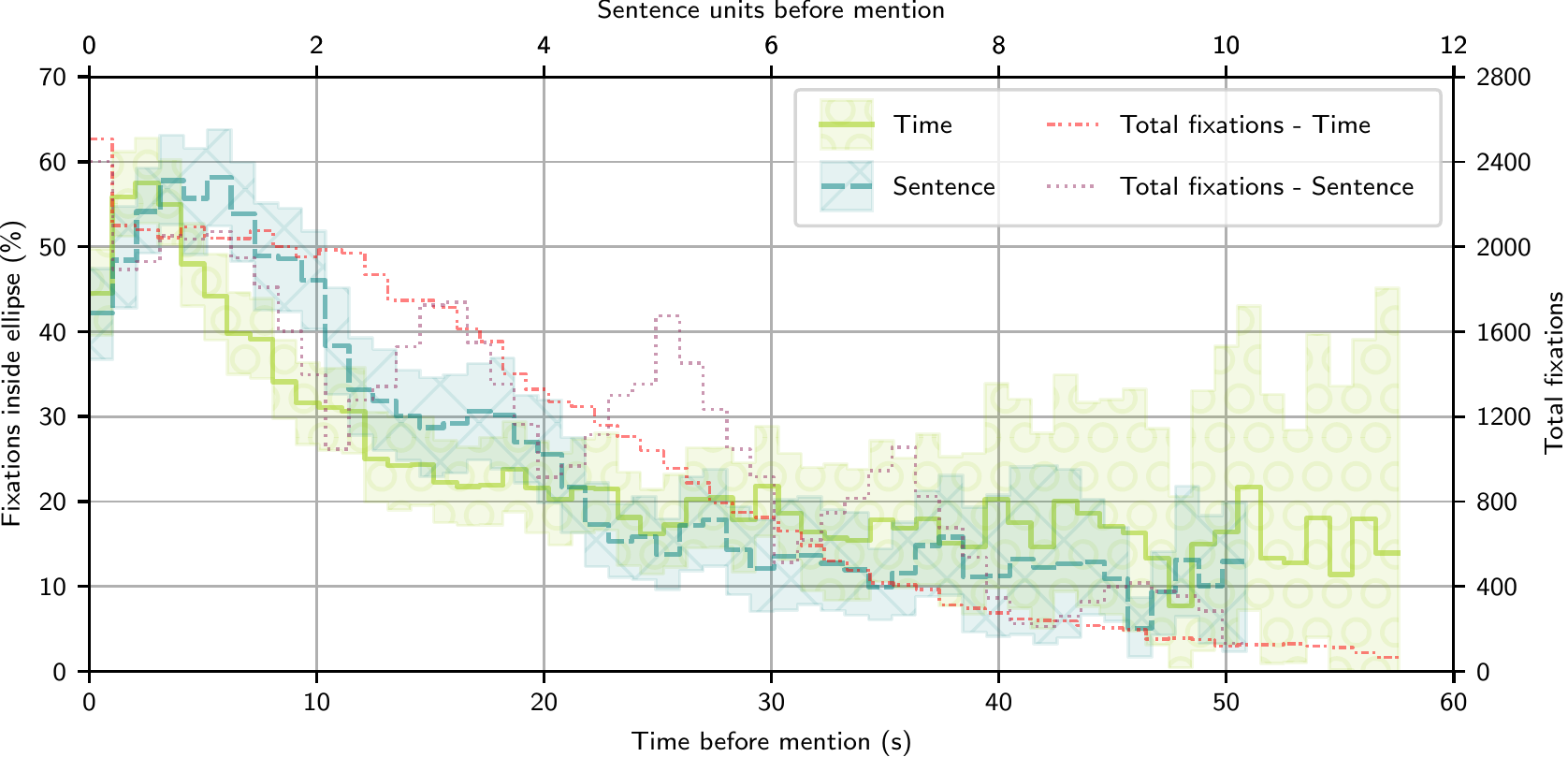}
\caption{Time analysis of the correlation between each mention of a label and what percentage of fixations were located inside the ellipses that localized each respective label. We present two lines, one as a function of time and another as a function of the counting of sentences and pauses before the mention. The step lines represent the percentage for separate data bins. We also draw the 95\% confidence interval for each bin in each line, calculated with bootstrapping. The number of fixations used to calculate each bin is shown in separate lines.}
\label{fig:delay}
\end{figure}

To analyze if the localization information given by the eye-tracking data correlates with the time that the presence of an abnormality was mentioned during the dictation, we produced the graph shown in Figure \ref{fig:delay}. To develop this analysis, we annotated the location where labels were mentioned in the report for 200 non-test \glspl*{CXR} from Phase 3. For annotating, we used a mix of a modified version of the chexpert labeler~\cite{chexpert} and hand-labeled corrections. \Glspl*{CXR} that had image-level labels not mentioned in the dictation were not included in the 200 randomly selected \glspl*{CXR}. 

We separated the time before the end of mentions of a label into bins of same size. For each bin, we calculated the percentage of fixations localized inside an ellipse for the same abnormality and \gls*{CXR}. 
The percentage was calculated considering the duration of the fixation inside the bin. Besides analyzing the delay in time units, we also used sentences units. To calculate the sentence units of each timestamp, we separated the dictation into mid-sentence moments and in-between-sentences pause moments. The timestamps of each transition between these were represented by integer numbers. Timestamps in the middle of a sentence or pause had their representation calculated through linear interpolation of the start and end of their sentence or pause. For example, a 12 s timestamp within the second sentence, dictated from 10 s to 15 s, would be associated with 3.4 sentence units. The sentence units shown in Figure \ref{fig:delay} represent the difference between the sentence units of the mentions and the fixations before the mentions. 

We divided the full range of data into 75 bins, of width 1.03 s and  0.21 sentences, and only kept bins up to before the first bin with less than or equal to ten fixations inside ellipses. We calculated 95 \% confidence intervals through bootstrapping, randomly sampling with replacement 200 \glspl*{CXR} from the 200 annotated \glspl*{CXR}. We performed this sampling 800 times and display in Figure \ref{fig:delay} the 2.5\% and 97.5\% percentiles for each bin.

As shown in Figure \ref{fig:delay}, there are peaks of correlation between the location of ellipses and the radiologists' fixations at around 2.5 s before the mention of the respective abnormality. The correlation peak was also calculated to be around 0.6 to 1.25 sentence units before the mention. This correlation shows that the transcription timestamps could be used to get label-specific localization information. With the shown delay between label fixation and mention, our data might need alignment algorithms for the correct association between fixations and label. We leave the exploration of the application of such algorithms to future works.

\subsection*{Validation labels}
\subsubsection*{Image-level labels}
The inter-rater reliability was measured through Fleiss' Kappa~\cite{fleiss} for the image-level labels, calculated using the statsmodels library in Python~\cite{statsmodels}. Image-level labels were considered positive when the maximum certainty, among all ellipses of a given label, was ``Possibly'' or higher. 
The inter-rater reliability scores for phases 1 and 2 are shown in Table~\ref{tab:fleiss}. The achieved inter-rater reliability scores are relatively low but in line with scores obtained in other similar studies for readings of \gls*{CXR}~\cite{lowrater1,lowrater2}. Some of the low values might be caused by the low prevalence of some of the labels~\cite{prevalence1,prevalence2}.
\subsubsection*{Localization ellipses}
For each \gls*{CXR}, for each label with more than one radiologist selecting certainty ``Possibly'' or higher, we calculated the average paired \gls*{IoU} between all pairs of respective radiologists. We then calculated the average \gls*{IoU} over all \glspl*{CXR} of each phase, with results presented in Table~\ref{tab:fleiss}. Between phases 1 and 2, we discussed a few examples of \gls*{CXR} that had low \gls*{IoU}.
\subsubsection*{Chest bounding boxes}
Similar to the localization ellipses, the quality of the bounding boxes containing the heart and the lungs was measured through \gls*{IoU}. \gls*{IoU} was calculated between all the pairs of radiologists for every \gls*{CXR} of the preliminary phases. For phase 1, the average \gls*{IoU} was 0.917$\pm$0.004 (n=59), and for phase 2, it was 0.920$\pm$0.004 (n=50). We organized no discussion for improvement of this label between phases 1 and 2.
\section*{Usage Notes}
To have access to the \glspl*{CXR} that the radiologists read, access to the MIMIC-CXR dataset~\cite{mimiccxr,mimiccxrdicom,mimiccxrjpg,physionet} is necessary. Both datasets are accessible only on Physionet, requiring the signature of a data use agreement by a logged-in user. Access to the MIMIC-CXR dataset requires free online courses in HIPAA regulations and human research subject protection.

The main uses intended for our data include:
\begin{itemize}
\item combining fixations into heatmaps, for use as an attention label and research on saliency maps and related subjects;
\item using the fixations as a nonuniformly sampled sequence of attention locations;
\item combining the timestamped transcriptions with the fixations for more specific localization to each abnormality;
\item associating the pupil data with the fixations for more information on the cognitive load of each fixation;
\item validating abnormalities parsed from the transcriptions using the image-level labels;
\item validating the locations found from the eye-tracking data through the abnormality ellipses;
\item using the chest bounding boxes for normalizing the location of the lungs while performing other analyses; and
\item using the chest bounding boxes for training a model to output bounding boxes for unseen data.
\end{itemize}
Other possible uses of the data may include using the certainties provided by the radiologists in uncertainty quantification research and the reports and their transcriptions in image captioning for chest x-rays. In \url{https://github.com/ricbl/eyetracking}, we provide examples on how to generate heatmaps, how to normalize the location of heatmaps using the chest bounding box, how to filter the fixations, how to calculate brightness of the shown \gls*{CXR} in any given time during dictation, and how to load and use the tables from the dataset.

\subsection*{Eye-tracking data}
There are uncertainties in the eye-tracking measurement pipeline. To represent them, we suggest following a method used in the generation of heatmaps in the visual attention modeling literature and modeling the location of each fixation as a Gaussian with a standard deviation of 1\degree\ of visual angle~\cite{onedegree}. We provide pixel resolution per visual angle for each axis of the image, so the Gaussian will be slightly anisotropic in the image space. For some applications, e.g., when generating one embedding vector per fixation for sequence analysis, it might be beneficial to filter out fixations that happened outside of the image. Among other reasons, fixations may have happened outside of the image because there were two buttons in the dictation screen: one to indicate that the dictation was over, and one to reset the windowing, zooming, and panning of the image. 

\subsection*{Abnormality labels}
\begin{figure}[ht]
\centering
\includegraphics[width=\textwidth]{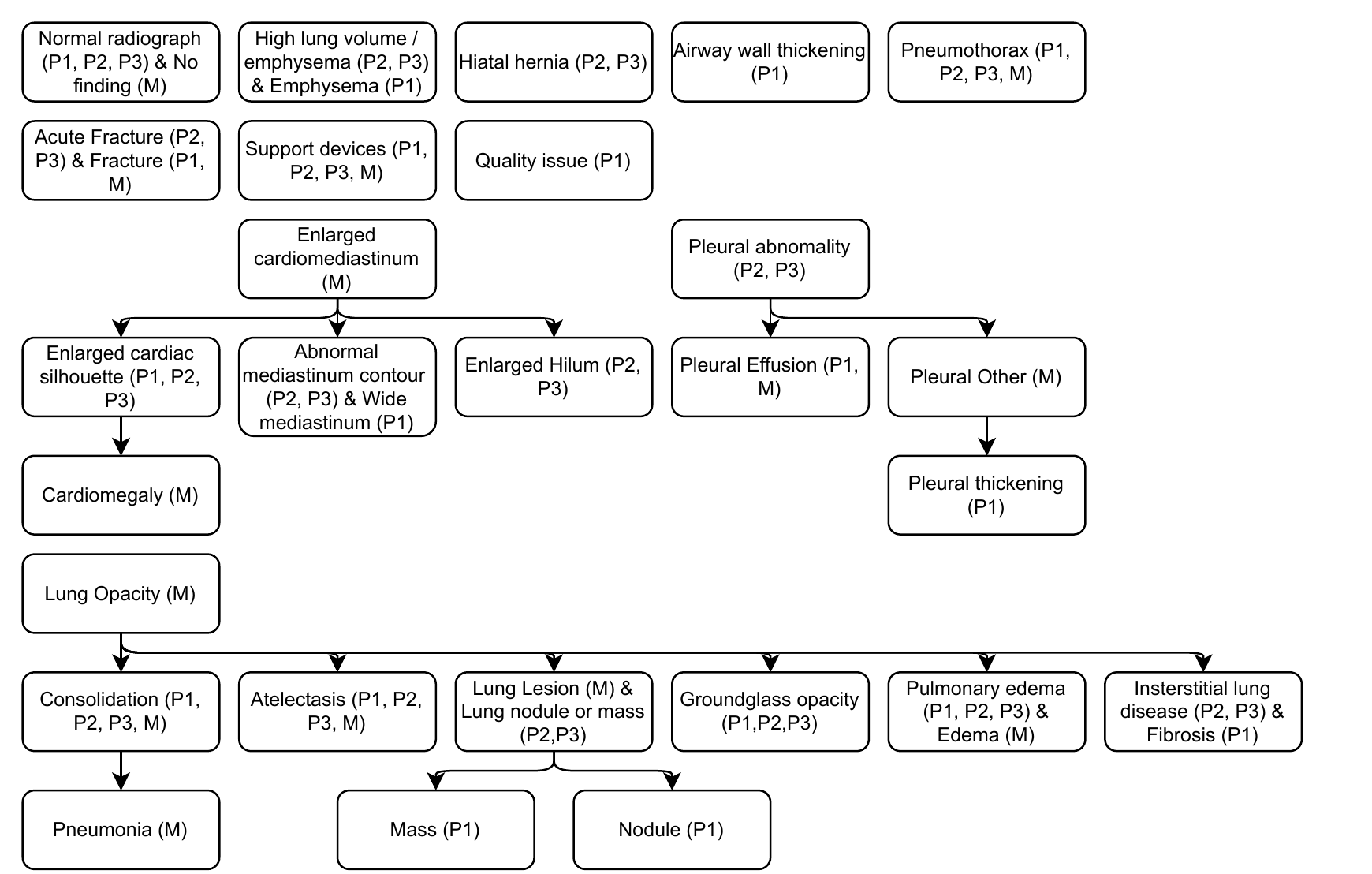}
\caption{Hierarchy of the labels of all the phases of our dataset and the labels of the MIMIC-CXR dataset. Arrows point to a subset of the originating label. The datasets to which each label belongs are listed inside parentheses, according to P1 (Phase 1), P2 (Phase 2), P3 (Phase 3), and M (MIMIC-CXR). Labels that do not have a hierarchical relationship with other labels are not connected to any arrows.}
\label{fig:labels}
\end{figure}

Figure~\ref{fig:labels} shows the hierarchy between the labels of our study and the labels from the MIMIC-CXR dataset. Not all labels present in one dataset have an equivalent in other datasets. This hierarchy was produced to the best of their understanding of the MIMIC-CXR labels. Supplementary File 1 provides the definition of the labels agreed upon among the radiologists who participated in the study.

\subsection*{Pupil data}

Bruny\'{e} et al.~\cite{pupilbrightness} showed that the pupil diameter of pathologists reliably increased for more difficult cases, providing an indicator of cognitive engagement. In our dataset, we provide the normalized pupil area, whose square root is equivalent to the normalized pupil diameter. The normalization was performed by a division by the area of the pupil in a standardized screen. However, the variation of the screen brightness, caused by windowing, zooming, and panning, may cause more variation in pupil area than psychological reasons. We suggest including another normalization, using the division by a value representing the screen brightness at each moment, similarly to what was done by Bruny\'{e} et al.~\cite{pupilbrightness}. For the calculation of this value, we suggest summing the intensity of pixels of the shown \gls*{CXR}. The part of the image shown in each moment and the part of the screen where it was shown are provided in the dataset. When considering the windowing of the \gls*{CXR}, images are shown following 
\begin{equation}
shown\_image = min\left(max\left(\frac{original\_image-window\_level}{window\_width}+0.5,0\right),1\right),
\end{equation}
where $window\_width$ can have values from 1.5e-05 to 2 and is usually initialized to 1, $window\_level$ can have values from 0 to 1 and is usually initialized to 0.5, $shown\_image$ is the image sent to the screen, and $original\_image$ is the loaded DICOM image normalized so that its possible range is from 0 to 1, which usually means dividing the image intensities by 4,096. The initial $window\_width$, $window\_level$, and the maximum intensity value were loaded from the DICOM tags of each image file.

\subsection*{Limitations of the study}
\begin{itemize}
\item Limited information presented to radiologists: our setup used a single screen, but multiple screens are used in a clinical setting. With multiple screens, the eye-tracker setup is more complex and would have to be validated. We limited the study to show only frontal \gls*{CXR}. Lateral views, past \glspl*{CXR}, and clinical information were not presented.
\item Report: in clinical practice, reports can be modified after a first transcription. We limited the editing to corrections of the transcription of the original dictation and deletions of dictation mistakes. This limitation was needed to assign timestamps to each word and ensure the radiologist saw the finding while the eye tracker was on. Several radiologists use templates for their reports in clinical practice, only dictating small parts of the report. We did not test such a dictation method. 
\item Head position: even though we used the remote mode for the eye tracker, which allows for some freedom of movement, the head movement, posture, and the distance from head to the screen were still more limited than in clinical practice, when radiologists can get closer to the screen to see a detail, for example. Because of this limitation, the use of zooming was probably more frequent than in clinical practice. Furthermore, radiologists mentioned that, with the limitations in position, they became fatigued faster than usual.
\item \gls*{CXR} dataset: we collected readings for images of only one dataset, so the current dataset may have ensuing biases. The radiologists also characterized images from the MIMIC-CXR dataset as having lower quality than usual for their practice, in aspects like the field-of-view excluding small parts of the lung and the blurring present in some images.
\item Display: the GPU of the computer used to display the \glspl*{CXR} supported only 8-bit display, so not all intensities of the original DICOM were shown, reducing the image quality and possibly changing the way radiologists interacted with the \gls*{CXR}. This limitation was partially remediated by allowing the windowing of the image to be changed.
\item Calibration cost: calibrations happened every 45 to 60 minutes, and sometimes more than five retries were needed to reach quality thresholds. The clinical implementation of the data collection method described in this paper for the collection of larges quantities of data might cause an undesired cost to the radiologist reading process.
\item Unautomated processes: another person was in the room coordinating calibrations and checking for low data quality. This same person raised the need for a recalibration or a change in position of the radiologist. This person might have to be replaced by automated processes if this data collection method is implemented in clinical practice.
\end{itemize}

\section*{Code availability}
The code used for all automatic processes described in this paper, involving sampling, collection, processing, and validation of data, is available at \url{https://github.com/ricbl/eyetracking}~\cite{mycodezenodo}. The software and versions we used were: MATLAB R2019a, Psychtoolbox 3.0.17~\cite{ptb1,ptb2,ptb3}, Python 3.7.7, edfapi 3.1, EYELINK II CL v5.15, Eyelink GL Version 1.2 Sensor=AC7, EDF2ASC 3.1, librosa 0.8.0~\cite{librosa}, numpy 1.19.1~\cite{numpy}, pandas 1.1.1~\cite{pandas,pandaszenodo}, matplotlib 3.5.1~\cite{matplotlib1,matplotlib2}, statsmodels 0.12.2~\cite{statsmodels}, shapely 1.7.1~\cite{shapely}, scikit-image 0.17.2~\cite{scikit-image}, pyrubberband 0.3.0, pydicom 2.1.2~\cite{pydicom}, pydub 0.24.1, soundfile 0.10.3.post1, pyttsx3 2.90, pillow 8.0.1~\cite{pillow}, scikit-learn 0.23.2~\cite{scikit-learn}, nltk 3.5~\cite{nltk}, syllables 1.0.0, moviepy 1.0.3~\cite{moviepy}, opencv 3.4.2~\cite{opencv_library}, Ubuntu 18.04.5 LTS, espeak 1.48.04, joblib 1.1.0, ffmpeg 3.4.8, and rubberband-cli 1.8.1.

\bibliography{main}

\section*{Acknowledgements}
This research was funded by the National Institute Of Biomedical Imaging And Bioengineering of the National Institutes of Health under Award Number R21EB028367. Christine Prickett provided copyediting support, Lauren Williams provided help in using Psychtoolbox 3, David Alonso provided advice in running eye-tracking sessions, Howard Mann was one of the five radiologists to read images, and Yichu Zhou was part of the team that modified the chexpert labeler to fit the labels of our dataset.

\section*{Author contributions statement}
R.B.L. wrote the manuscript, did all the coding specific for this project, conducted the data-collection sessions, and ran the analyses. 
M.Z. participated in the design of the technical validation analysis. 
W.A. participated in the study design, was one of the readers and provided feedback for data-collection processes. 
J.C. was one of the readers and provided feedback for data-collection processes. 
P.A.D. was one of the readers and provided feedback for data-collection processes. 
V.S. participated in the study design. 
T.D. participated in the study design, conceived the fundamental parts of the eye-tracking side of the data-collection sessions, and wrote the paragraph comparing types of eye-tracking devices. 
J.S. participated in the study design,  guided the clinical decisions for the data collection, produced the clinical instructions, was one of the readers, and provided feedback for data-collection processes. 
T.T. is the PI of the project, coordinating the study design, leading discussions about the project, and editing the manuscript.
All authors reviewed the manuscript.

\section*{Competing interests}

The authors declare no competing interests.

\section*{Figures \& Tables}

\begin{figure}[ht]
\centering
\includegraphics[width=\textwidth]{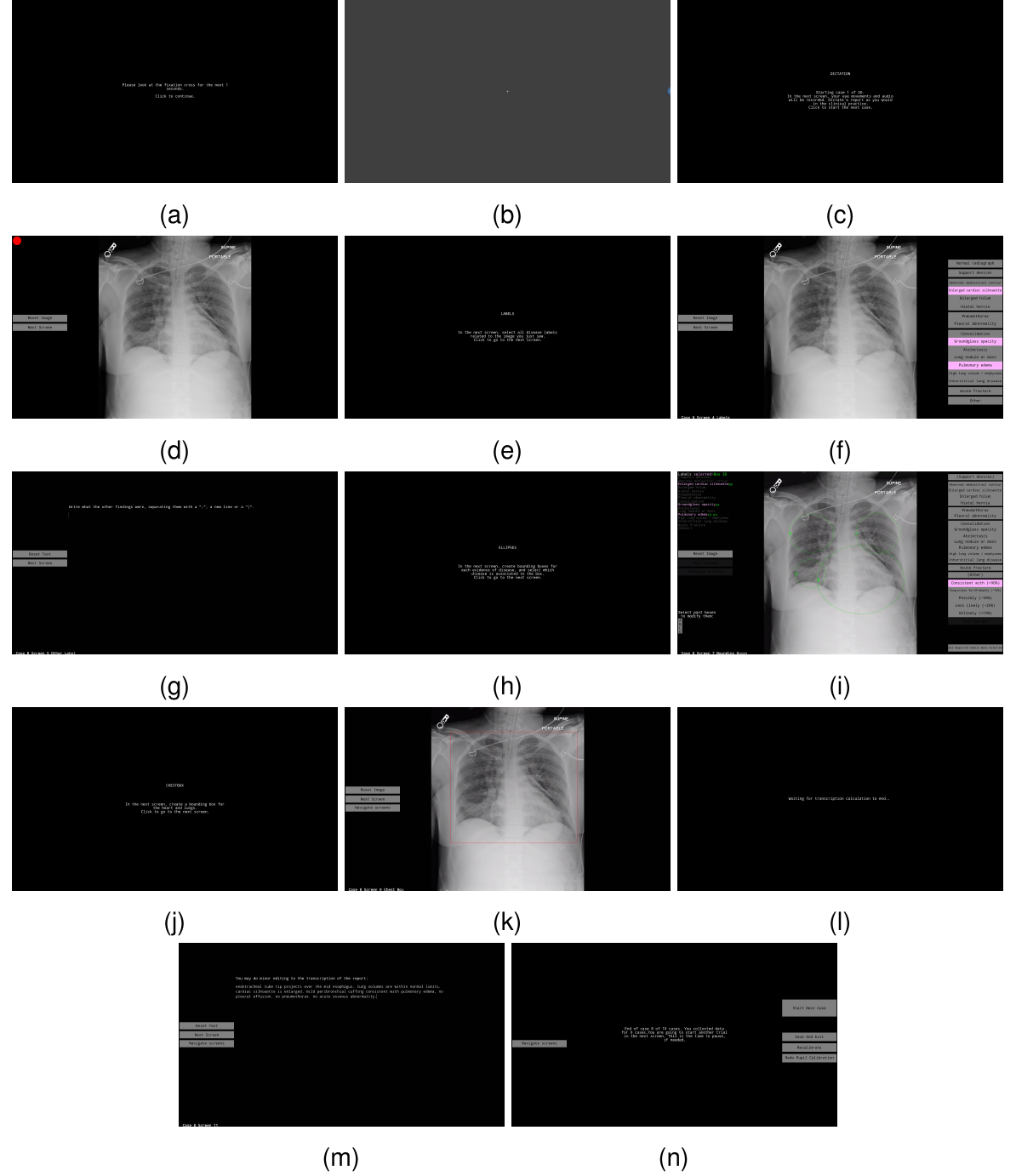}
\caption{Screens of the data-collection interface in the sequence they are presented to a radiologist, including instruction screens (a, c, e, h, j, n), calibration of pupil size (b), dictation of reports (d), choice of global labels (f, g), selection of ellipses and certainties (i),  drawing of lung/heart box (k), and editing of transcription (l, m). Digital visualization is recommended for reading the content.}
\label{fig:screens}
\end{figure}

\begin{table}[ht]
\centering
\begin{tabular}{|l|r|r|r|r|}
\hline
Label                                                                                                                                     & FK P1 & FK P2    & IoU P1   & IoU P2 \\ \hline
Airway wall thickening (P1)                                                                                                                 & 0.03$\pm$0.20  &           & \makecell[r]{0.28\quad\quad\,\,
\\(n=4)}     &        \\ \hline
Atelectasis (P1, P2)                                                                                                                              & 0.34$\pm$0.05  & 0.25$\pm$0.08      & \makecell[r]{0.29$\pm$0.03\\(n=33)}      & \makecell[r]{0.37$\pm$0.04\\(n=19)}   \\ \hline
Consolidation (P1, P2)                                                                                                                             & 0.36$\pm$0.07  & 0.50$\pm$0.07      & \makecell[r]{0.36$\pm$0.04\\(n=25)}      & \makecell[r]{0.37$\pm$0.04\\(n=17)}   \\ \hline
\makecell[l]{Emphysema (P1)\\\& High lung volume / emphysema (P2)} & 0.26$\pm$0.32  & 0.10$\pm$0.34      & \makecell[r]{0.43\quad\quad\,\,
\\(n=3)}      & \makecell[r]{0.53\quad\quad\,\,
\\(n=2)}   \\ \hline
Enlarged cardiac silhouette (P1, P2)                                                                                                              & 0.56$\pm$0.07  & 0.55$\pm$0.08      & \makecell[r]{0.75$\pm$0.02\\(n=20)}      & \makecell[r]{0.75$\pm$0.02\\(n=20)}   \\ \hline
Enlarged hilum (P2)                                                                                                                           &       & -0.03$\pm$0.36     &           &    Undefined    \\ \hline
Fibrosis (P1) \& Interstitial lung disease (P2)  & 0.29$\pm$0.44  & 0.16$\pm$0.57      & \makecell[r]{0.53\quad\quad\,\,\\(n=1)}       & \makecell[r]{0.23\quad\quad\,\,\\(n=1)}   \\ \hline
Fracture (P1) \& Acute fracture (P2)   & 0.25$\pm$0.25  & 0.71$\pm$0.36      & \makecell[r]{0.42\quad\quad\,\,
\\(n=4)}     & \makecell[r]{0.21\quad\quad\,\,\\(n=1)}   \\ \hline
Groundglass opacity (P1, P2)                                                                                                                      & 0.10$\pm$0.17  & 0.21$\pm$0.11      & \makecell[r]{0.42\quad\quad\,\,
\\(n=6)}      & \makecell[r]{0.38$\pm$0.05\\(n=16)}   \\ \hline
Hiatal hernia  (P2)                                                                                                                           &       & Undefined &           &    Undefined    \\ \hline
Mass (P1)                                                                                                                                     & -0.01$\pm$0.70 &           & Undefined &        \\ \hline
Nodule (P1)                                                                                                                                   & 0.29$\pm$0.25  &           & \makecell[r]{0.38\quad\quad\,\,
\\(n=2)}      &        \\ \hline
Lung nodule or mass (P2)                                                                                                                      &       & 0.36$\pm$0.49      &           & \makecell[r]{0.83\quad\quad\,\,\\(n=1)}   \\ \hline
Pleural abnormality (P2)                                                                                                                     &       & 0.67$\pm$0.07      &           & \makecell[r]{0.27$\pm$0.02\\(n=18)}   \\ \hline
Pleural effusion (P1)                                                                                                                         & 0.53$\pm$0.06  &           & \makecell[r]{0.37$\pm$0.03\\(n=22)}       &        \\ \hline
Pleural thickening (P1)                                                                                                                       & 0.06$\pm$0.40  &           & \makecell[r]{0.29\quad\quad\,\,\\(n=1)}      &        \\ \hline
Pneumothorax (P1, P2)                                                                                                                             & 0.55$\pm$0.25  & 0.62$\pm$0.28      & \makecell[r]{0.41\quad\quad\,\,
\\(n=3)}      & \makecell[r]{0.27\quad\quad\,\,
\\(n=3)}    \\ \hline
Pulmonary edema (P1, P2)                                                                                                                          & 0.22$\pm$0.13  & 0.10$\pm$0.14      & \makecell[r]{0.25$\pm$0.03\\(n=11)}      & \makecell[r]{0.36$\pm$0.06\\(n=11)}   \\ \hline
Quality issue (P1)                                                                                                                           & 0.02$\pm$0.30  &           &  &        \\ \hline
Support devices (P1, P2)                                                                                                                         & 0.77$\pm$0.05  & 0.47$\pm$0.06      &  &        \\ \hline
\makecell[l]{Wide mediastinum (P1)\\\& Abnormal mediastinal contour (P2)} & 0.23$\pm$0.34  & 0.05$\pm$0.25      & \makecell[r]{0.57\quad\quad\,\,
\\(n=2)}      & \makecell[r]{0.58\quad\quad\,\,
\\(n=3)}   \\ \hline
\end{tabular}
\caption{\label{tab:fleiss} Inter-rater scores for validation of the quality of the data. For phases 1 (P1) and 2 (P2), we present reliability on image-level labels, calculated using Fleiss' Kappa (FK), and average IoU of the abnormality ellipses. All scores are paired with standard errors. The number of samples given for the IoU values represents the number of independent CXRs used in the calculation. The phases in which each label was present is listed in parenthesis. Table cells are left blank for labels that were not present in a specific phase.}
\end{table}

\begin{table}[ht]
\centering
\begin{tabular}{|l|r|r|r|r|}
										\hline
Dataset	&	\makecell[c]{Phase 1\\(P1)}	&	\makecell[c]{Phase 2\\(P2)}	&	\makecell[c]{Phase 3\\(P3)}	&	\makecell[c]{MIMIC-CXR filtered\\(M)}	\\	\hline
\# cases	&	295	&	250	&	2,507	&	194,495	\\	\hline
\# cases studies with eye-tracking data	&	285	&	240	&	2,507	&	0	\\	\hline
\# MIMIC-CXR images	&	59	&	50	&	2,507	&		\\	\hline
\# subjects & 58 & 50 & 2,110 & 60,018 \\ \hline
\% female	&	63.8	&	54.0	&	50.7	&	53.9	\\	\hline
\% male	&	36.2	&	46.0	&	49.1	&	45.7	\\	\hline
\% test set	&	15.3	&	14.0	&	20.2	&	1.4	\\	\hline
\% Normal Radiograph (P1, P2, P3) \& No Finding (M)	&	18.0	&	24.4	&	22.8	&	32.9	\\	\hline
\makecell[l]{\% Abnormal mediastinal contour (P2,P3) \&\\Wide mediastinum (P1)}	&	2.7	&	5.6	&	2.7	&		\\	\hline
\% Acute fracture (P2,P3) \& Fracture (P1, M)	&	5.1	&	2.8	&	1.0	&	1.9	\\	\hline
\% Airway wall thickening (P1)	&	7.1	&		&		&		\\	\hline
\% Atelectasis (P1,P2,P3,M)	&	41.4	&	27.6	&	25.8	&	20.5	\\	\hline
\% Cardiomegaly (M)	&		&		&		&	19.8	\\	\hline
\% Consolidation (P1,P2,P3,M)	&	28.5	&	28.8	&	25.9	&	4.7	\\	\hline
\% Enlarged cardiac silhouette (P1,P2,P3)	&	28.1	&	28.4	&	21.8	&		\\	\hline
\% Enlarged Cardiomediastinum (M)	&		&		&		&	3.2	\\	\hline
\% Enlarged hilum (P2,P3)	&		&	2.8	&	1.9	&		\\	\hline
\% Groundglass opacity (P1,P2,P3)	&	9.2	&	18.8	&	12.6	&		\\	\hline
\% Hiatal hernia (P2,P3)	&		&	0.0	&	0.9	&		\\	\hline
\makecell[l]{\% High lung volume / emphysema (P2,P3) \&\\Emphysema (P1)}	&	3.1	&	3.2	&	2.9	&		\\	\hline
\% Interstitial lung disease (P2,P3) \& Fibrosis (P1)	&	1.7	&	1.2	&	1.0	&		\\	\hline
\% Lung nodule or mass (P2,P3) \& Lung Lesion (M)	&		&	1.6	&	5.1	&	2.7	\\	\hline
\% Lung Opacity (M)	&		&		&		&	22.8	\\	\hline
\% Mass (P1)	&	0.7	&		&		&		\\	\hline
\% Nodule (P1)	&	4.7	&		&		&		\\	\hline
\% Other (P1,P2,P3)	&	13.9	&	8.8	&	6.0	&		\\	\hline
\% Pleural abnormality (P2,P3)	&		&	30.0	&	29.5	&		\\	\hline
\% Pleural Effusion (P1,M)	&	31.2	&		&		&	24.2	\\	\hline
\% Pleural thickening (P1)	&	2.0	&		&		&		\\	\hline
\% Pleural Other (M)	&		&		&		&	0.9	\\	\hline
\% Pneumonia (M)	&		&		&		&	7.2	\\	\hline
\% Pneumothorax (P1,P2,P3,M)	&	4.7	&	4.4	&	2.9	&	4.6	\\	\hline
\% Pulmonary edema (P1,P2,P3) \& Edema (M)	&	13.9	&	13.6	&	13.7	&	12.1	\\	\hline
\% Quality issue (P1)	&	3.4	&		&		&		\\	\hline
\% Support devices (P1,P2,P3,M)	&	36.9	&	34.8	&	44.8	&	29.3	\\	\hline

\end{tabular}
\caption{\label{tab:stats} Statistics of each phase of data collection and the subset of the MIMIC-CXR dataset from which images were sampled. The dataset where each label was present is shown inside parentheses. ``Normal radiograph'' represents CXRs for which no other label was selected. Table cells are left blank for labels that were not present in that dataset. For how the labels of the different datasets are related, check Figure \ref{fig:labels}.}
\end{table}

\end{document}